\documentclass[11pt]{article}
\usepackage[utf8]{inputenc}      
\usepackage[T1]{fontenc}         
\usepackage{amsmath, amssymb}    
\usepackage{amsfonts, mathrsfs}  
\usepackage{amsthm}              
\usepackage{bm}                  
\usepackage{graphicx}            
\usepackage{color}               
\usepackage{cite}                
\usepackage{enumerate}           
\usepackage{geometry}            
\usepackage{float}               
\usepackage{ifthen}              
\usepackage[caption=false]{subfig} 
\usepackage{epstopdf}            
\usepackage{braket}              
\usepackage{hyperref}            

\begin{document}

\title{\bf Uncertainty relations for unified ($\alpha$,$\beta$)-relative entropy of coherence under mutually unbiased equiangular tight frames}

\vskip0.1in
\author{\small Baolong Cheng$^1$, Zhaoqi Wu$^1$\thanks{Corresponding author. E-mail: wuzhaoqi\_conquer@163.com}\\
{\small\it  1. Department of Mathematics, Nanchang University,
Nanchang 330031, P R China} }

\date{}
\maketitle

\noindent {\bf Abstract} {\small }\\
Uncertainty relations based on quantum coherence is an important
problem in quantum information science. Uncertainty relations for
unified ($\alpha$,$\beta$)-relative entropy of coherence under
mutually unbiased equiangular tight frames. We discuss uncertainty
relations for averaged unified ($\alpha$,$\beta$)-relative entropy
of coherence under mutually unbiased equiangular tight frames, and
derive an interesting result for different parameters. As
consequences, we obtain corresponding results under mutually
unbiased bases, equiangular tight frames or based on Tsallis
$\alpha$- relative entropies and R\'enyi-$\alpha$ relative
entropies. We illustrate the derived inequalities by explicit
examples in two dimensional spaces, showing that the lower bounds
can be regarded as good approximations to averaged coherence
quantifiers under certain circumstances.

\vskip0.2in

\noindent {\bf 1. Introduction}\par

Uncertainty relations, lying at the heart of quantum mechanics,
remain one of the core issues in quantum information science.
The first uncertainty relation was proposed by
Heisenberg \cite{Heisenberg}, while Robertson gave the lower bound
of the product of the variances of two
observables\cite{RobertsonHP}. Thereafter, uncertainty relations
based on the Shannon entropy of the measurement outcomes were
further proposed by Deutsch \cite{Deutsch}, Maassen and Uffink
\cite{Maassen}, and the latter is state-independent. Berta et al.
\cite{Berta} proposed the uncertainty relations using the
conditional entropy, while the majorization entropic uncertainty
relation \cite{Puchala} and strong majorization entropic uncertainty
relation \cite{Rudnicki} were considered based on the R\'enyi
entropy and Tsallis entropy. The entropic uncertainty relations has
many applications in quantum information theory, such as quantum
random number generation \cite{Vallone}, quantum metrology
\cite{Giovannetti}, and quantum teleportation \cite{HuML}.

Studies of nonclassical correlations in quantum
information processing constitute an essential part of recent
developments. A quantification scheme for coherence resource was
introduced \cite{Baumgratz} and an equivalent framework has been
proposed \cite{YuX} which maybe easier to verify in certain
situations. The study on quantum coherence from the perspective of
resource theory has attracted widespread attention\cite{Streltsov}.
For the classical information entropy, the unified
($\alpha$,$\beta$)-entropy \cite{RathiePN} and two-parameter
generalization of the R\'enyi entropy \cite{GhoshA1} were introduced
as extensions of R\'enyi entropy, while the unified
($\alpha$,$\beta$)-relative entropy\cite{WangJ}, generalized
relative ($\alpha$,$\beta$)-entropy\cite{GhoshA2} and generalized
alpha-beta divergence\cite{GhoshA3} have been proposed respectively.
The quantum unified ($\alpha$,$\beta$) entropy was introduced in
\cite{HuX}, while the quantum unified ($\alpha$,$\beta$)-relative
entropy was put forward in \cite{WangJ}, which are generalizations
of quantum R\'enyi $\alpha$ relative entropy\cite{Mosonyi2011} and
quantum Tsallis $\alpha$ relative
entropy\cite{AbeS200301,AbeS200302}. As important coherence
quantifiers, the R\'enyi $\alpha$-relative entropy of coherence
\cite{ShaoL} and the Tsallis $\alpha$-relative entropy of coherence
\cite{Rastegin2016,ZhaoH} were proposed, respectively, while the
unified ($\alpha$,$\beta$)-relative entropy of coherence has been
defined and its analytical formulas has been deduced \cite{MuH}.

The selection of different computational bases depending on the
theoretical and experimental context. Mutually unbiased bases (MUBs) was first discussed in \cite{Schwinger}, and its properties with prime dimension has been investigated by Ivonovic\cite{Ivonovic}. Two observables are so-called complementary if their eigenvectors are mutually unbiased in finite dimensions\cite{Kraus}, and exact knowledge of the measured value of one observable means maximal uncertainty in the other. Although the existence of $d+1$ MUBs for $d$ being a prime power has been proved \cite{Durt}, its existence in general dimensions is unsolved. The MUBs are applied in some popular schemes of quantum cryptography due to the fact that detecting a particular basis state reveals no information about the state, which was prepared in another basis \cite{Rastegin2013}. They have also been used in the BB84 scheme of quantum key distribution \cite{Bennett}, entanglement detection \cite{Spengler2012,ShangJ}, and the quantum error correction codes \cite{Spengler2013}. Symmetric informationally complete measurements (SIC-POVMs) are closely related with MUBs and have a lot of common applications \cite{Beneduci,Bengtsson,Renes}.

Equiangular tight frames (ETFs, Optimal Grassmannian frames), which
can induce a SIC-POVM under certain conditions, have specific
properties on finite-dimensional spaces, and have important
applications in wireless communication and multiple description
coding\cite{Strohmer}. Finite tight frames \cite{Waldron}, as a
natural generalization of orthonormal bases, are useful in many
areas like coding and signal processing \cite{Casazza}. ETFs yield
an optimal packing of lines in a Euclidean space, and can be applied
to build a positive operator-valued measurements. Furthermore, the
concepts of MUBs and ETFs have been extended to the one of mutually
unbiased equiangular tight frames (MUETFs) \cite{Fickus}.

The complementarity relations for quantum coherence under a complete
set of mutually unbiased bases (the upper bounds of coherence) were
first proposed in \cite{ChengS}, and various forms of
coherence-mixedness tradeoffs were addressed \cite{ZhangQH}. On the
other hand, uncertainty relations for coherence based on SIC-POVMs
\cite{LuoSL} and MUBs \cite{LuoSL,ShenMY,ShengYH,ZhangFG} have been
studied extensively. In particular, the uncertainty relations for
the relative entropy of coherence with respect to MUBs
\cite{Rastegin2018}, Tsallis $\frac{1}{2}$-relative entropy of
coherence under MUBs and ETFs \cite{Rastegin2022}, and Tsallis
$\alpha$-relative entropy of coherence under MUBs and ETFs
\cite{Rastegin20241} (the lower bounds of coherence) have been
discussed, respectively. Recently, the uncertainty relations for
Tsallis $\alpha$-relative entropy of coherence under MUETFs have
been derived\cite{Rastegin20242}. In this paper, we explore
uncertainty relations for coherence quantifiers via unified
($\alpha$,$\beta$)-relative entropy under measurements assigned to
MUETFs.

The remainder of this paper is structured as follows. In Section 2,
we recall some preliminary concepts. The main results and some
corollaries are presented in Section 3. We also exemplify the
derived inequalities with SIC-POVMs and MUBs in Section 4. Some
concluding remarks are given in Section 5.

\vskip0.1in

\noindent {\bf 2. Preliminaries}\par In this section, we recall the
definitions of both classical and quantum information entropies, the
framework of coherence and the coherence quantifiers we will use in
this paper, the concepts of mutually unbiased equiangular tight
frames and related ones. Throughout this paper, we denote by
$\mathbb{R}$ the set of real numbers, $\mathbb{R}^+$ the set of
positive real numbers, $\mathbb{Z^{+}}$ the set of positive integers
and $\mathbb{N}$ the set of natural numbers, i.e.,
$\mathbb{N}=\mathbb{Z^{+}}\cup \{0\}$.

\noindent{\bf 2.1 Unified ($\alpha$,$\beta$) entropy and unified
($\alpha$,$\beta$)-relative entropy}\par

In this subsection, we discuss the relations between unified
($\alpha$,$\beta$) entropy (unified ($\alpha$,$\beta$)-relative
entropy) and some other two parameter generalizations of classical
entropies in corresponding literatures. For the sake of convenience,
let the space of probability distributions over a finite alphabet
set $\{a_1,a_2,\cdots,a_n\}$ be
$$
\Omega_n = \left\{P=(p_1,p_2,\cdots,p_n): p_i=\mathrm{Prob}(a_i)\ge0, \text{ for all } i=1,\cdots,n, W(P):=\sum_ip_i=1\right\},
$$
and the set of finite sub-probability distributions be
$$
\Omega_n^* = \left\{P=(p_1,p_2,\cdots,p_n):
p_i=\mathrm{Prob}(a_i)\ge0, \text{ for all } i=1,\cdots,n,
W(P):=\sum_ip_i\le1\right\}.
$$

For any $\alpha\in\mathbb{R}^+$, $\beta\in\mathbb{R}$, and
$P\in\Omega_n$, the unified ($\alpha$,$\beta$) entropy is defined
as\cite{RathiePN}
\begin{align}
E_{\alpha}^{\beta}(P) =
\begin{cases}
H_{\alpha}^{\beta}(P), & \text{if $\alpha\ne1$, $\beta\ne0$},\\
H_{\alpha}(P), & \text{if $\alpha\ne1$, $\beta=0$},\\
H^{\alpha}(P), & \text{if $\alpha\ne1$, $\beta=1$},\\
_{\frac{1}{\alpha}}H(P),& \text{if $\alpha\ne1$, $\beta=\frac{1}{\alpha}$},\\
H(P), & \text{if $\alpha=1$,}
\label{eq1}
\end{cases}
\end{align}
where
\begin{align}
\begin{aligned}
H_{\alpha}^{\beta}(P) =& \frac{1}{(1-\alpha)\beta}\left[\left(\sum_{i=1}^{n}p_i^\alpha\right)^\beta-1\right],\\
H_{\alpha}(P) =& \frac{1}{1-\alpha}\mathrm{ln}\left(\sum_{i=1}^{n}p_i^\alpha\right),\\
H^{\alpha}(P) =& \frac{1}{1-\alpha}\left(\sum_{i=1}^{n}p_i^\alpha-1\right),\\
_{\frac{1}{\alpha}}H(P) =& \frac{1}{\alpha-1}\left[\left(\sum_{i=1}^{n}p_i^{\frac{1}{\alpha}}\right)^\alpha-1\right],\\H(P) =& -\sum_{i=1}^{n}p_i\mathrm{ln}p_i.
\notag
\end{aligned}
\end{align}
For any $\alpha,\beta\in\mathbb{R}^+$, a two-parameter
generalization of the R\'enyi entropy of $P\in\Omega_n^*$ is defined
as \cite{GhoshA1}
\begin{align}
\begin{aligned}
\mathcal{E}_{\alpha,\beta}^{LN}(P):=\frac{\alpha\beta}{\alpha-\beta}\mathrm{ln}\left[\frac{\left(\sum\limits_{i=1}^{n}p_i^\beta\right)^{\frac{1}{\beta}}}{\left(\sum
\limits_{i=1}^{n}p_i^\alpha\right)^{\frac{1}{\alpha}}}\right],\text{ }\alpha\ne\beta.
\label{eq2}
\end{aligned}
\end{align}
Note that when $\alpha,\beta\in\mathbb{R}^+$, $\alpha\ne1$,
$\alpha\ne\beta$ and $P\in\Omega_n$, the quantities in \eqref{eq1}
and \eqref{eq2} exhibit the following relation
\begin{align}
\begin{aligned}
\mathcal{E}_{\alpha,\beta}^{LN}(P)=\frac{1}{\alpha-\beta}\mathrm{ln}\left[\frac{(1-\beta)\alpha H_\beta^\alpha(P)+1}{(1-\alpha)\beta H_\alpha^\beta(P)+1}\right]=\frac{\alpha\beta}{\alpha-\beta}\mathrm{ln}\left[\frac{(\frac{1}{\alpha}-1) H_\alpha^{\frac{1}{\alpha}}(P)+1}{(\frac{1}{\beta}-1) H_\beta^{\frac{1}{\beta}}(P)+1}\right].
\label{eq3}
\end{aligned}
\end{align}
Furthermore, letting $\alpha\to\beta$, we obtain
\begin{align}
\begin{aligned}
\lim_{\alpha \to \beta} \mathcal{E}_{\alpha,\beta}^{LN}(P)=\frac{1}{\beta}\mathrm{ln}[(1-\beta)\beta H_\beta^\beta(P)+1]+\beta\mathcal{E}_{\beta}^{AD}(P),
\label{eq4}
\end{aligned}
\end{align}
where
$\mathcal{E}_{\beta}^{AD}(P)=-\frac{\sum\limits_{i=1}^{n}p_i^\beta\mathrm{ln}p_i}{\sum\limits_{i=1}^{n}p_i^\beta}$
is the so-called Aczel-Daroczy entropy \cite{AczelJ}.

Note that $H_{\alpha}^{\beta}(P)$ defined on a probability
distribution and $\mathcal{E}_{\alpha,\beta}^{LN}(P)$ defined on a
sub-probability distribution are both nonnegative, continuous and
concave ($H_{\alpha}^{\beta}(P)$ is concave for $0<\alpha\le1$,
$\alpha\beta\le1$ or $\alpha\ge1$, $\alpha\beta\ge1$, while
$\mathcal{E}_{\alpha,\beta}^{LN}(P)$ is concave for $0<\beta\le1$,
$\alpha\ge\beta$ or $0<\alpha\le1$, $\beta\ge\alpha$). Both of them
have decisivity (i.e., the entropy functional satisfies
$\mathcal{E}(P)=0$ for $P=(0,1)$) and expandability (i.e.,
$\mathcal{E}(P)=\mathcal{E}((p_1,p_2,\cdots,p_n,0))$ for
$P=(p_1,p_2,\cdots,p_n)\in \Omega_n^*$). However, it can be seen
that $\mathcal{E}_{\alpha,\beta}^{LN}(P)$ is symmetric with respect
to $\alpha$ and $\beta$, while $H_{\alpha}^{\beta}(P)$ is not. None
of them satisfy the branching/recursivity property (i.e.,
$\mathcal{E}(p_1,p_2,\cdots,p_{n-1},p_nq_1,p_nq_2,\cdots,p_nq_m)=\mathcal{E}(p_1,p_2,\cdots,p_n)+p_n\mathcal{E}(q_1,\cdots,q_m)$
for any $P=(p_1,\cdots,p_n)\in\Omega_n$ and
$Q=(q_1,\cdots,q_m)\in\Omega_m$). Moreover, it holds that $
H_{\alpha}^{\beta}(P*Q)=H_{\alpha}^{\beta}(P)+H_{\alpha}^{\beta}(Q)+(1-\alpha)\beta
H_{\alpha}^{\beta}(P)H_{\alpha}^{\beta}(Q)$ for
$\alpha\in\mathbb{R}^+$ and $\beta\in\mathbb{R}$, yet $
\mathcal{E}_{\alpha,\beta}^{LN}(P*Q)=\mathcal{E}_{\alpha,\beta}^{LN}(P)+\mathcal{E}_{\alpha,\beta}^{LN}(Q)$
for $\alpha,\beta\in\mathbb{R}^+$, where
$P*Q=(p_iq_j)_{i=1,\cdots,n;j=1,\cdots,m}$ for $P\in\Omega_n$ and
$Q\in\Omega_m$.

For any $\alpha\in\mathbb{R}^+$, $\beta\in\mathbb{R}$, and
$P,Q\in\Omega_n$, the unified ($\alpha$,$\beta$)-relative entropy is
defined by \cite{WangJ}
\begin{align}
E_{\alpha}^{\beta}(P\parallel Q) =
\begin{cases}
H_{\alpha}^{\beta}(P\parallel Q), & \text{if $\alpha\ne1$, $\beta\ne0$},\\
H_{\alpha}(P\parallel Q), & \text{if $\alpha\ne1$, $\beta=0$},\\
H^{\alpha}(P\parallel Q), & \text{if $\alpha\ne1$, $\beta=1$},\\
_{\frac{1}{\alpha}}H(P\parallel Q),& \text{if $\alpha\ne1$, $\beta=\frac{1}{\alpha}$},\\
H(P\parallel Q), & \text{if $\alpha=1$,}
\label{eq5}
\end{cases}
\end{align}
where
\begin{align}
\begin{aligned}
H_{\alpha}^{\beta}(P\parallel Q) =& \frac{1}{(\alpha-1)\beta}\left[\left(\sum_{i=1}^{n}p_i\frac{p_i^{\alpha-1}}{q_i^{\alpha-1}}\right)^\beta-1\right],\alpha>0,\\
H_{\alpha}(P\parallel Q) =& \frac{1}{1-\alpha}\mathrm{ln}\left(\sum_{i=1}^{n}p_i\frac{p_i^{\alpha-1}}{q_i^{\alpha-1}}\right),\alpha>0,\\
H^{\alpha}(P\parallel Q) =& \frac{1}{1-\alpha}\left(\sum_{i=1}^{n}p_i\frac{p_i^{\alpha-1}}{q_i^{\alpha-1}}-1\right),\alpha>0,\\
_{\frac{1}{\alpha}}H(P\parallel Q) =& \frac{1}{\alpha-1}\left[\left(\sum_{i=1}^{n}p_i\frac{p_i^{\frac{1}{\alpha}-1}}{q_i^{\frac{1}{\alpha}-1}}\right)^\alpha-1\right],\alpha>0,\\H(P\parallel Q) =& -\sum_{i=1}^{n}p_i\mathrm{ln}\frac{p_i}{q_i}.
\notag
\end{aligned}
\end{align}

For any $\alpha\in\mathbb{R}^+$ and any $\beta\in\mathbb{R}$, put
$\lambda=\frac{\beta}{\alpha}-1$. Suppose that $P$, $Q$ are two
probability distributions on a measurable space and have absolutely
continuous densities $p$ and $q$, respectively, with respect to a
common dominating $\sigma$-finite measure $\mu$. Then the relative
($\alpha$,$\beta$)-entropy is defined as \cite{GhoshA2}
\begin{align}
\begin{aligned}
\mathcal{R}\mathcal{E}_{\alpha,\beta}(P,Q)=\frac{1}{\beta\lambda}\mathrm{log}[\mathrm{sign}(\beta\lambda)D_\lambda(P_\alpha,Q_\alpha)]+1,
\label{eq6}
\end{aligned}
\end{align}
where $D_\lambda(P_\alpha,Q_\alpha)=\frac{1}{\lambda-1}\mathrm{log}\left(\int p_\alpha^\lambda q_\alpha^{1-\lambda} \mathrm{d}\mu\right)$, and $P_\alpha$, $Q_\alpha$ are defined by
$$
\frac{\mathrm{d}P_\alpha}{\mathrm{d}\mu} = p_\alpha = \frac{p^\alpha}{\int p^\alpha \, \mathrm{d}\mu}, \quad \frac{\mathrm{d}Q_\alpha}{\mathrm{d}\mu} = q_\alpha = \frac{q^\alpha}{\int q^\alpha \, \mathrm{d}\mu}.
$$

Another generalized divergence was defined based on a class of
generating functions. Let $\psi:[0,\infty]\to\mathbb{R}$ be a
suitable transformation, for any $\alpha\beta(\alpha+\beta)\ne0$,
the generalized alpha-beta divergence between two sub-probability
distributions $P$ and $Q$ is defined as \cite{GhoshA3}
\begin{align}
\begin{aligned}
d_{GAB}^{(\alpha,\beta),\psi}(P,Q)=\frac{1}{\beta(\alpha+\beta)}\psi\left(\left \| p \right \|_{\alpha+\beta}^{\alpha+\beta} \right)+\frac{1}{\alpha(\alpha+\beta)}\psi\left(\left \| q \right \|_{\alpha+\beta}^{\alpha+\beta} \right)-\frac{1}{\alpha\beta}\psi\left(\left \langle p,q  \right \rangle _{\alpha,\beta}\right),
\label{eq7}
\end{aligned}
\end{align}
where $\left \| p \right \|_{\alpha+\beta}=\int p^{\alpha+\beta} \,
\mathrm{d}\mu$, $\left \| q \right \|_{\alpha+\beta}=\int
q^{\alpha+\beta} \, \mathrm{d}\mu$ and $\left \langle p,q  \right
\rangle_{\alpha,\beta}=\int p^\alpha q^\beta \, \mathrm{d}\mu$.

{\bf Remark 1} Note that $H_{\alpha}^{\beta}(P\parallel Q)$ is a
distance measure between two probability distributions for discrete
random variables, while
$\mathcal{R}\mathcal{E}_{\alpha,\beta}(P,Q)$/$d_{GAB}^{(\alpha,\beta),\psi}(P,Q)$
are distance measures between two probability
distributions/sub-probability distributions for continuous random
variables. Moreover, $H_{\alpha}^{\beta}(P\parallel Q)$ reduces to
the Tsallis $\alpha$-relative entropy, the R\'enyi $\alpha$-relative
entropy and the relative entropy respectively, when $\beta=1$,
$\beta\to0$, and $\alpha\to1$, and
$\mathcal{R}\mathcal{E}_{\alpha,\beta}(P,Q)$ reduces to the scaled
R\'enyi $\beta$-relative entropy when $\alpha=1$. Also, for
$\beta=1-\alpha$, $\alpha\notin \{0,1\}$ and $P,Q\in \Omega_n$,
$d_{GAB}^{(\alpha,\beta),\psi}(P,Q)$ reduces to the scaled Tsallis
$\alpha$-relative entropy and the scaled R\'enyi $\alpha$-relative
entropy respectively, when $\psi(x)=x$ and $\psi(x)=\mathrm{ln}x$.

\noindent{\bf 2.2 Coherence quantifiers of the quantum unified
($\alpha$,$\beta$)-relative entropy}\par Let $\mathcal{H}$ be a
$d$-dimensional Hilbert space, $\mathcal{A}=\{\ket{i}\}_{i=1}^d$ a
reference basis of $\mathcal{H}$, and $\mathcal{D(H)}$ the set of
density matrices (quantum states) on $\mathcal{H}$. The set of
incoherent states is defined by \cite{Baumgratz}
$$
\mathcal{I} = \left\{\sigma\in \mathcal{D(H)}\Bigg
|\sigma=\sum_{i=1}^{d}\sigma_i\ket{i}\bra{i}\right\},
$$
i.e., an incoherent state is a quantum state which are diagonal
under the given basis.

$C(\rho)$ is called a coherence measure of the quantum state $\rho$,
if $C(\cdot)$ satisfies the following conditions\cite{Baumgratz}:

(1) nonnegativity: $C(\rho)\ge0$ and $C(\rho)=0$ iff $\rho\in\mathcal{I}$;

(2) monotonicity: $C(\rho)\ge C(\Phi(\rho))$, where $\Phi$ is any incoherent completely positive and trace-preserving map;

(3) strong monotonicity: $\sum\limits_n p_n C(\rho_n)\le C(\rho)$, where $p_n = \mathrm{tr}(K_n \rho K_n^\dagger)$ and $\rho_n = \frac{K_n \rho K_n^\dagger}{\mathrm{tr}(K_n \rho K_n^\dagger)}$ for all $K_n$ with $\sum\limits_n K_n^\dagger K_n = I$ and $K_n \mathcal{I} K_n^\dagger \subseteq \mathcal{I}$;

(4) convexity: $C\left(\sum\limits_i p_i \rho_i\right) \le \sum\limits_i p_i C(\rho_i)$ for any ensemble $\{ p_i,\rho_i \}$.

For any $\alpha \in$ [0,1] and $\beta\in\mathbb{R}$, the unified
($\alpha$,$\beta$)-relative entropy is defined by \cite{WangJ}
\begin{align}
D_{\alpha}^{\beta}(\rho\parallel\sigma) =
\begin{cases}
H_{\alpha}^{\beta}(\rho\parallel\sigma), & \text{if $0\le\alpha<1$, $\beta\ne0$},\\
H_{\alpha}(\rho\parallel\sigma), & \text{if $0\le\alpha<1$, $\beta=0$},\\
H^{\alpha}(\rho\parallel\sigma), & \text{if $0\le\alpha<1$, $\beta=1$},\\
_{\frac{1}{\alpha}}H(\rho\parallel\sigma),& \text{if $0<\alpha<1$, $\beta=\frac{1}{\alpha}$},\\
H(\rho\parallel\sigma), & \text{if $\alpha=1$,}
\label{eq8}
\end{cases}
\end{align}
where
\begin{align}
\begin{aligned}
H_{\alpha}^{\beta}(\rho\parallel\sigma) =& \frac{1}{(\alpha-1)\beta}[(\mathrm{tr}(\rho^{\alpha}\sigma^{1-\alpha}))^\beta-1],\\
H_{\alpha}(\rho\parallel\sigma) =& \frac{1}{\alpha-1}\mathrm{ln}(\mathrm{tr}(\rho^{\alpha}\sigma^{1-\alpha})),\\
H^{\alpha}(\rho\parallel\sigma) =& \frac{1}{\alpha-1}[\mathrm{tr}(\rho^{\alpha}\sigma^{1-\alpha})-1],\\
_{\frac{1}{\alpha}}H(\rho\parallel\sigma) =& \frac{1}{1-\alpha}[(\mathrm{tr}(\rho^{\frac{1}{\alpha}}\sigma^{1-\frac{1}{\alpha}}))^\alpha-1],\\H(\rho\parallel\sigma) =& \mathrm{tr}(\rho\mathrm{ln}\rho)-\mathrm{tr}(\rho\mathrm{ln}\sigma).
\label{eq9}
\end{aligned}
\end{align}

{\bf Remark 2} Note that $H_{\alpha}^{\beta}(\rho\parallel\sigma)$
reduces to the quantum Tsallis $\alpha$-relative entropy, the quantum R\'enyi
$\alpha$-relative entropy and the quantum relative entropy respectively,
when $\beta=1$, $\beta\to0$, and $\alpha\to1$.

For any $\alpha \in (0,1)$ and $\beta\le1$, the unified
($\alpha$,$\beta$)-relative entropy of coherence (UREOC) \cite{MuH}
is defined as
\begin{align}\label{eq10}
C_{(\alpha,\beta)}(\mathcal{A};\rho) = \min_{\sigma\in\mathcal{I}}D_{\alpha}^{\beta}(\rho\parallel\sigma).
\end{align}

It has been proved that $C_{(\alpha,\beta)}(\mathcal{A};\rho)$ is a
coherence monotone \cite{MuH}, and its analytical formula is
expressed as \cite{MuH}
\begin{align}\label{eq11}
C_{(\alpha,\beta)}(\mathcal{A};\rho) = \dfrac{1}{(\alpha-1)\beta}\left[\left(\sum_{i=1}^{d}\bra{i}{\rho}^{\alpha}{\ket{i}}^{\frac{1}{\alpha}}\right)^{\alpha\beta}-1\right].
\end{align}

{\bf Remark 3} $C_{(\alpha,\beta)}(\mathcal{A};\rho)$ reduces to the
Tsallis $\alpha$-relative entropy of coherence
$C_\alpha(\mathcal{A};\rho)$ and the R\'enyi $\alpha$-relative
entropy of coherence $\widetilde{C}_\alpha(\mathcal{A};\rho)$
respectively, when $\beta=1$ and $\beta\to0$.

\noindent{\bf 2.3 Mutually unbiased equiangular tight frames}\par
Let $\mathcal{H}$ be a $d$-dimensional Hilbert space. Two orthonormal bases $\mathcal{B}_1 = \{\ket{j_1}\}$ and $\mathcal{B}_2 = \{\ket{j_2}\}$ in $\mathcal{H}$ are said to be mutually unbiased \cite{Schwinger}, if for all $j_1$ and $j_2$,
\begin{align}\label{eq12}
\left | \left \langle j_1| j_2 \right \rangle \right | = \dfrac{1}{\sqrt[]{d} }.
\end{align}

When $d$ is a prime power, i.e. $d = p^M$ where $p$ is prime number and $M$ is constant, there exist sets of $d + 1$ MUBs, and these sets are maximal in the sense that it is impossible to find more than $d + 1$ MUBs in any $\mathcal{H}$ \cite{Durt}.

The set $\mathbb{B} = \{\mathcal{B}_1, \mathcal{B}_2, \cdots, \mathcal{B}_M\}$ is called a set of mutually unbiased bases (MUBs), when each two terms of $\mathbb{B}$ are mutually unbiased. We are interested in this strong condition which can help us to improve entropic uncertainty relations \cite{Coles}. If two observables have unbiased eigenbases, then the measurement of one observable reflect no information about possible outcomes of the measurement of others, so the states in MUBs are indistinguishable in this sense \cite{Durt}.

In the following, we will consider only complex frames. A set of unit vectors $\{\ket{\varphi_j}\}_{j=1}^{N}$ $(N\ge d)$ is called a frame \cite{Casazza}, if for all unit vector $\ket{\psi}\in\mathcal{H}$, there exists $0<S_0<S_1<\infty$ such that
\begin{align}\label{eq13}
S_0 \le \sum_{j=1}^{N}{\left | \left \langle \varphi_j| \psi \right \rangle \right |}^2 \le S_1,
\end{align}
where $S_0$ and $S_1$ are the minimal and maximal eigenvalues of the frame operator $\sum\limits_{j=1}^{N}\ket{\varphi_j}\bra{\varphi_j}$, respectively.

Furthermore, the frame is called a tight frame \cite{Sustik} in the case that $S_0 = S_1 = S$ with $S = \dfrac{N}{d}$. Moreover, the tight frame is called equiangular \cite{Strohmer}, if for $N\le d^2$, it holds that
\begin{align}\label{eq14}
{\left | \left \langle \varphi_i| \varphi_j \right \rangle \right |}^2 = \dfrac{N-d}{d(N-1)}\text{\quad}(i\ne j).
\end{align}
It is obvious that a Parseval tight frame obtained by setting $S=1$
is equivalent to a set of orthonormal bases. Based on any ETF, we
can construct the POVM $\mathcal{P}$ as
\begin{align}\label{eq15}
\mathcal{P} = \Bigg\{P_j\Bigg| P_j = \dfrac{d}{N}\ket{{\varphi}_j}\bra{{\varphi}_j}\text{, } \sum_{j=1}^{N}P_j = \mathbb{I}_d\Bigg\}.
\end{align}
When the measured state is described by a quantum state $\rho$ with $\mathrm{tr}\rho = 1$, the probability of $j$-th outcome is given by
\begin{align}\label{eq16}
p_j(P_j;\rho) = \dfrac{d}{N}\bra{\varphi_j}\rho\ket{\varphi_j}.
\end{align}

When $N=d^2$, \eqref{eq14} becomes
\begin{align}\label{eq17}
\left | \left \langle \varphi_i | \varphi_j \right \rangle \right | = \dfrac{1}{\sqrt{d+1}}  \quad(i\ne j).
\end{align}
In this case, $\{\ket{\varphi_j}\}_{j=1}^{N}$ induces a SIC-POVM \cite{Renes}
\begin{align}\label{eq18}
\mathcal{F} = \Bigg\{F_j\Bigg| F_j = \dfrac{1}{d}\ket{{\varphi}_j}\bra{{\varphi}_j}\text{, } \sum_{j=1}^{d}F_j = \mathbb{I}_d\Bigg\},
\end{align}
in which $\{F_j\}$ is a set of $d^2$ rank-one operators on
$\mathcal{H}$. There are indications that SIC-POVMs exist in all
dimensions. However, although many explicit constructions for
SIC-POVMs have been given, a universal method still lacks.
Therefore, we prefer to employ ETFs which may be easier to construct
than SIC-POVMs.

Suppose that $M\ge1$. A set of unit vectors $\{\ket{\varphi_{\mu,j}}\}$ with $\mu=1,\cdots,M$ and $j=1,\cdots,N$ forms a MUETF \cite{Fickus} if
\begin{align}
{\left | \left \langle \varphi_{\mu,i}| \varphi_{v,j} \right \rangle \right |}^2 =
\begin{cases}
c, & \text{if $\mu=v$ and $i\ne j$},\\
\frac{1}{d}, & \text{if $\mu\ne v$},
\label{eq19}
\end{cases}
\end{align}
where $c=\dfrac{N-d}{d(N-1)}$. It is obvious that a MUETF consists
of $M$ usual mutually unbiased ETFs, so it reduce to an ETF when
$M=1$ and MUBs when $N=d$ and $c=0$. Each MUETF induces a set of
POVMs
\begin{align}\label{eq20}
\mathcal{F_\mu}=\Bigg\{F_{\mu,j}\Bigg| F_{\mu,j}=\frac{d}{N}\ket{\varphi_{\mu,j}}\bra{\varphi_{\mu,j}}\text{, } \sum_{j=1}^{N}F_{\mu,j} = \mathbb{I}_d\Bigg\}.
\end{align}

We can assign a nonorthogonal resolution of the identity to each of $M$ ETFs with the probabilities
\begin{align}\label{eq21}
p_j(\mathcal{F}_\mu;\rho) = \dfrac{d}{N}\bra{\varphi_{\mu,j}}\rho\ket{\varphi_{\mu,j}},
\end{align}
where the corresponding index of coincidence reads as
$$
I(\mathcal{F}_\mu;\rho) = \sum\limits_{j=1}^{d}p_j^2(\mathcal{F}_\mu;\rho).
$$

The coherence quantifier $C_{(\alpha,\beta)}(\mathcal{A};\rho)$ in
\eqref{eq11} under $\mathcal{F}_\mu$ can be written as
\begin{align}\label{eq22}
C_{(\alpha,\beta)}(\mathcal{F}_\mu;\rho) = \dfrac{1}{(\alpha-1)\beta}\left\{\left[\sum_{j=1}^{N}\left(\frac{d}{N}\bra{\varphi_{\mu,j}}{\rho}^{\alpha}{\ket{\varphi_{\mu,j}}}\right)^{\frac{1}{\alpha}}\right]^{\alpha\beta}-1\right\}.
\end{align}

{\bf Remark 4} In the same manner,
$C_{(\alpha,\beta)}(\mathcal{F}_\mu;\rho)$ reduces to the Tsallis
$\alpha$-relative entropy of coherence
$C_\alpha(\mathcal{F}_\mu;\rho)$ and the R\'enyi $\alpha$-relative
entropy of coherence $\widetilde{C}_\alpha(\mathcal{F}_\mu;\rho)$
respectively, when $\beta=1$ and $\beta\to0$.

\vskip0.1in

\noindent {\bf 3. Uncertainty relations of UREOC under MUETFs}\par
In this section, we first present the uncertainty relations of UREOC
under MUETFs. We then obtain a series of corollaries corresponding
to the degradation of MUETFs to MUBs and ETFs, and UREOC to the
coherence quantifiers based on Tsallis $\alpha$-relative entropy and
R\'enyi $\alpha$-relative entropy.

Let us begin with the $\gamma$-logarithm of positive variable defined as
\begin{align}
\mathrm{ln}_\gamma(X) =
\begin{cases}
\frac{X^{1-\gamma}-1}{1-\gamma}, & \text{if $0\le\gamma\ne1$},\\
\mathrm{ln}(X), & \text{if $\gamma=1$}.
\label{eq23}
\end{cases}
\end{align}

For $\gamma\in\mathbb{R}^+$, the Tsallis $\gamma$-entropy \cite{Tsallis} reads as
$$
H_\gamma(P)=\frac{1}{1-\gamma}\left(\sum\limits_{j=1}^Np_j^\gamma-1\right)=\sum\limits_{j=1}^Np_j\mathrm{ln}_\gamma\left(\frac{1}{p_j}\right).
$$

Suppose that $k\in \mathbb{Z^{+}}$. We define the piecewise smooth function as
$$
L_\gamma(X)=(k+1)\mathrm{ln}_\gamma(k+1)-k\mathrm{ln}_\gamma(k)-k(k+1)[\mathrm{ln}_\gamma(k+1)-\mathrm{ln}_\gamma(k)]X, X\in\left[\frac{1}{k+1},\frac{1}{k}\right].
$$

To prove the main results, we first present the following three lemmas.

{\bf Lemma 1} \cite{Rastegin20231} For any $\gamma\in(0,2]$, we have
$$
H_\gamma(P)\ge L_\gamma(I(P)),
$$
where
$$
I(P) = \sum\limits_{j=1}^{N}p^2_j.
$$

{\bf Lemma 2} \cite{Rastegin20242} For a MUETF with the corresponding index of coincidence, we have
$$
\frac{1}{M}\sum\limits_{\mu=1}^{M}I(\mathcal{F}_\mu;\rho)\le\frac{(1-c)[d\mathrm{tr}\rho^2-1]}{MNS}+\frac{1}{N},
$$
where $c=\dfrac{N-d}{d(N-1)}$.

{\bf Lemma 3} Suppose that $x\ge0$ and $l\in \mathbb{Z^{+}}$. For any $\alpha\in(0,1)$ and $\beta\in[0,1]$, define a piecewise linear function as
$$
L_{(\alpha,\beta)}(x) = f(l) + \frac{f(l+1)-f(l)}{(l+1)-l}(x-l), x\in[l,l+1],
$$
where $f(x) = \frac{x}{1-(\alpha-1)\beta x}$. Then it holds that
$$
f(x)\ge L_{(\alpha,\beta)}(x)\ge0,x\in[l,l+1].
$$

\textit{Proof} It is easy to calculate that
$$
f(x)\ge0,f^{'}(x) = \frac{1}{(1-ax)^2}>0,f^{''}(x) = \frac{2a}{(1-ax)^3}<0
$$
for $x\ge0$ and $a=(\alpha-1)\beta\le0$. Thus, $f(x)$ is strictly increasing and concave. Based on the properties of $f(x)$, it is obvious that $L_{(\alpha,\beta)}(x)$ is increasing and is a chord of $f(x)$. Thus we have $f(x)\ge L_{(\alpha,\beta)}(x)\ge0$ for $x\in[l,l+1]$. This completes the proof.\qed

{\bf Remark 5}  $f(x) = L_{(\alpha,\beta)}(x)$ if $x\in\mathbb{N}$ or $\beta = 0$.

We are now ready to give our main results.

{\bf Theorem 1} Let $\{\ket{\varphi_{\mu,j}}\}$ with
$\mu=1,\cdots,M$ and $j=1,\cdots,N$ be a MUETF in $\mathcal{H}$,
where $N\ge d$, and $\rho\in\mathcal{D(H)}$. Then we have

(1) For any $\alpha\in[\frac{1}{2},1)$ and
$\beta\in(-\infty,0)\cup(0,1]$, it holds that
\begin{align}
\frac{1}{M}\sum\limits_{\mu=1}^{M}C_{(\alpha,\beta)}(\mathcal{F}_\mu;\rho)\ge&\frac{(\mathrm{tr}\rho^\alpha)^\beta}{(\alpha-1)\beta}\Bigg\{\frac{\alpha-1}{\alpha}\mathrm{L}_{\frac{1}{\alpha}}\Bigg(\frac{(1-c)[d\mathrm{tr}\rho^{2\alpha}(\mathrm{tr}\rho^\alpha)^{-2}-1]}{MNS}\notag\\
&+\frac{1}{N}\Bigg)+1\Bigg\}^{\alpha\beta}-\frac{1}{(\alpha-1)\beta};
\label{eq24}
\end{align}

(2) For any $\alpha\in(0,\frac{1}{2})$ and $\beta\in(-\infty,0)$, it
holds that
\begin{align}
\frac{1}{M}\sum\limits_{\mu=1}^{M}C_{(\alpha,\beta)}(\mathcal{F}_\mu;\rho)\ge&-\frac{(\mathrm{tr}\rho^{1-\alpha})^{\beta}}{\alpha\beta}\Bigg\{\frac{\alpha}{\alpha-1}\mathrm{L}_{\frac{1}{1-\alpha}}\Bigg(\frac{(1-c)[d\mathrm{tr}\rho^{2(1-\alpha)}(\mathrm{tr}\rho^{1-\alpha})^{-2}-1]}{MNS}\notag\\
&+\frac{1}{N}\Bigg)+1\Bigg\}^{(1-\alpha)\beta}+\frac{1}{\alpha\beta};
\label{eq25}
\end{align}

(3) For any $\alpha\in(0,\frac{1}{2})$ and $\beta\in(0,1]$, it holds
that
\begin{align}
\frac{1}{M}\sum\limits_{\mu=1}^{M}C_{(\alpha,\beta)}(\mathcal{F}_\mu;\rho)\ge&L_{(\alpha,\beta)}\Bigg(\frac{(\mathrm{tr}\rho^{1-\alpha})^{-\beta}}{\alpha\beta}\Bigg\{\frac{\alpha}{\alpha-1}\mathrm{L}_{\frac{1}{1-\alpha}}\Bigg(\frac{(1-c)[d\mathrm{tr}\rho^{2(1-\alpha)}(\mathrm{tr}\rho^{1-\alpha})^{-2}-1]}{MNS}\notag\\
&+\frac{1}{N}\Bigg)+1\Bigg\}^{(\alpha-1)\beta}-\frac{1}{\alpha\beta}\Bigg),
\label{eq26}
\end{align}
where $\mathcal{F_\mu}$ are the induced POVMs given in \eqref{eq20}, $c=\dfrac{N-d}{d(N-1)}$ and $S=\frac{N}{d}$.

\textit{Proof} (1) For any $\alpha\in[\frac{1}{2},1)$, let $\gamma=\frac{1}{\alpha}$, then $\gamma\in(1,2]$. For the given state $\rho$, define
$$
\delta=\frac{\rho^\alpha}{\mathrm{tr}\rho^\alpha}.
$$
Then we have
$$
C_{(\alpha,\beta)}(\mathcal{F}_\mu;\rho)=\dfrac{\gamma}{(\gamma-1)\beta}-\dfrac{\gamma(\mathrm{tr}\rho^{\frac{1}{\gamma}})^\beta}{(\gamma-1)\beta}\left[\sum_{j=1}^{N}\left(\frac{d}{N}\bra{\varphi_{\mu,j}}\delta{\ket{\varphi_{\mu,j}}}\right)^\gamma\right]^{\frac{\beta}{\gamma}}.
$$

According to Lemma 1, we have
\begin{align}
\frac{1}{M}\sum\limits_{\mu=1}^{M}C_{(\alpha,\beta)}(\mathcal{F}_\mu;\rho) \ge \frac{\gamma}{(\gamma-1)\beta}-\dfrac{\gamma(\mathrm{tr}\rho^{\frac{1}{\gamma}})^\beta}{(\gamma-1)\beta}\sum_{\mu=1}^{M}\frac{1}{M}\left\{  1-(\gamma-1)L_\gamma[I(\mathcal{F}_\mu;\rho)] \right\}^{\frac{\beta}{\gamma}}.
\label{eq27}
\end{align}

\textbf{Case 1}. If $\beta\in(0,1]$, since $f:X\mapsto L_\gamma(X)$ is non-increasing and convex, and $g:Y\mapsto -\left[1-(\gamma-1)Y\right]^{\frac{\beta}{\gamma}}$ is non-decreasing and convex, it follows that the composition of them
$$
g\circ f:X \mapsto -\left[1-(\gamma-1)L_\gamma(X)\right]^{\frac{\beta}{\gamma}}
$$
is non-increasing and convex. Then we have
\begin{align}
\frac{1}{M}\sum\limits_{\mu=1}^{M}C_{(\alpha,\beta)}(\mathcal{F}_\mu;\rho) \ge \frac{\gamma}{(\gamma-1)\beta}-\dfrac{\gamma(\mathrm{tr}\rho^{\frac{1}{\gamma}})^\beta}{(\gamma-1)\beta}\left\{  1-(\gamma-1)L_\gamma\left[\sum_{\mu=1}^{M}\frac{I(\mathcal{F}_\mu;\rho)}{M}\right] \right\}^{\frac{\beta}{\gamma}}.
\label{eq28}
\end{align}

\textbf{Case 2}. If $\beta\in[-1,0)$, since $f:X\mapsto L_\gamma(X)$ is non-increasing and convex, and $g:Y\mapsto\frac{1}{\left[1-(\gamma-1)Y\right]^{-\frac{\beta}{\gamma}}}$ is non-decreasing and convex, it follows that the composition of them
$$
g\circ f:X \mapsto \frac{1}{\left[1-(\gamma-1)L_\gamma(X)\right]^{-\frac{\beta}{\gamma}}}
$$
is non-increasing and convex. Rewriting \eqref{eq27} as
$$
\frac{1}{M}\sum\limits_{\mu=1}^{M}C_{(\alpha,\beta)}(\mathcal{F}_\mu;\rho) \ge \dfrac{\gamma(\mathrm{tr}\rho^{\frac{1}{\gamma}})^\beta}{(1-\gamma)\beta}\sum_{\mu=1}^{M}\frac{1}{M}\frac{1}{\left\{  1-(\gamma-1)L_\gamma[I(\mathcal{F}_\mu;\rho)] \right\}^{-\frac{\beta}{\gamma}}}-\frac{\gamma}{(1-\gamma)\beta},
$$
we also obtain \eqref{eq28}.

\textbf{Case 3}. If $\beta\in(-\infty,-1)$, then $-\beta\in(1,+\infty)$. Since $f:X\mapsto \frac{1}{\left[1-(\gamma-1)L_\gamma(X)\right]^{\frac{1}{\gamma}}}$ is non-increasing and convex, and $g:Y\mapsto Y^{-\beta}$ non-decreasing and convex, it follows that the composition of them
$$
g\circ f:X \mapsto \Bigg\{\frac{1}{\left[1-(\gamma-1)L_\gamma(X)\right]^{\frac{1}{\gamma}}}\Bigg\}^{-\beta}
$$
is non-increasing and convex. Rewriting \eqref{eq27} as
$$
\frac{1}{M}\sum\limits_{\mu=1}^{M}C_{(\alpha,\beta)}(\mathcal{F}_\mu;\rho) \ge \dfrac{\gamma(\mathrm{tr}\rho^{\frac{1}{\gamma}})^\beta}{(1-\gamma)\beta}\sum_{\mu=1}^{M}\frac{1}{M}\left\{\frac{1}{\left\{  1-(\gamma-1)L_\gamma[I(\mathcal{F}_\mu;\rho)] \right\}^{\frac{1}{\gamma}}}\right\}^{-\beta}-\frac{\gamma}{(1-\gamma)\beta},
$$
\eqref{eq28} follows immediately. This implies that \eqref{eq28} holds in all cases. Combining \eqref{eq28} with Lemma 2, we obtain \eqref{eq24}. Therefore, item (1) holds.

(2) Since $\alpha\in(0,\frac{1}{2})$, we have $1-\alpha\in(\frac{1}{2},1)$. For $\beta\in(-\infty,0)$, according to Theorem 3.5(2) in \cite{WangJ}, we have
$$
C_{(\alpha,\beta)}(\mathcal{F}_\mu;\rho)\ge C_{(1-\alpha,\beta)}(\mathcal{F}_\mu;\rho).
$$
Substituting $\alpha$ by $1-\alpha$ in \eqref{eq24}, we then obtain \eqref{eq25}. So item (2) is proved.

(3) For any $\alpha\in(0,\frac{1}{2})$ and $\beta\in(0,1]$, we have $-\beta\in[-1,0)$. Accroding to Lemma 3, we have
$$
C_{(\alpha,\beta)}(\mathcal{F}_\mu;\rho) = \frac{C_{(\alpha,-\beta)}(\mathcal{F}_\mu;\rho)}{1-(\alpha-1)\beta C_{(\alpha,-\beta)}(\mathcal{F}_\mu;\rho)}\ge L_{(\alpha,\beta)}(C_{(\alpha,-\beta)}(\mathcal{F}_\mu;\rho)),
$$
which implies that
$$
\frac{1}{M}\sum\limits_{\mu=1}^{M}C_{(\alpha,\beta)}(\mathcal{F}_\mu;\rho) \ge L_{(\alpha,\beta)}\Bigg(\frac{1}{M}\sum\limits_{\mu=1}^{M}C_{(\alpha,-\beta)}(\mathcal{F}_\mu;\rho)\Bigg).
$$
Combining this with \eqref{eq25}, we obtain \eqref{eq26}. Hence we have derived item (3). This completes the proof.\qed

{\bf Remark 6} (1) We claim that the lower bounds in \eqref{eq24}-\eqref{eq26} are always nonnegative. In fact, for any $\alpha\in[\frac{1}{2},1)$ and $\beta\in(-\infty,0)\cup(0,1]$, denote the right hand side of \eqref{eq24} by $A(\alpha,\beta)$, $X=\frac{(1-c)[d\mathrm{tr}\rho^{2\alpha}(\mathrm{tr}\rho^\alpha)^{-2}-1]}{MNS}+\frac{1}{N}$, it is obvious that $\mathrm{L}_\gamma(X)\ge0$, and $\mathrm{L}_\gamma(X)=0$ iff $X=1$, which is equivalent to $\mathrm{tr}\rho^\alpha=\sqrt{\frac{d\mathrm{tr}\rho^{2\alpha}}{\frac{MS(N-1)}{1-c}+1}}$. Since $\alpha\in[\frac{1}{2},1)$, we have
\begin{align}
\mathrm{tr}\rho^\alpha&\le\sqrt{\frac{d}{\frac{M(N-1)^2}{d-1}+1}}\le\sqrt{\frac{d}{\frac{(d-1)^2}{d-1}+1}}=1.\notag
\end{align}
On the other hand, it holds that $\mathrm{tr}\rho^\alpha\ge1$ for all $\alpha\in[\frac{1}{2},1)$. This implies that when $\mathrm{tr}\rho^\alpha=1$, we have $L_\gamma(X)=0$, which yields that
$$
A(\alpha,\beta)\ge\frac{\left(\mathrm{tr}\rho^\alpha\right)^\beta-1}{(\alpha-1)\beta}=0.
$$

Since the right hand side of \eqref{eq25} can be obtained by substituting $\alpha$ by $1-\alpha$ in \eqref{eq24}, it is also nonnegative. Finally, since $L_{(\alpha,\beta)}(x)\ge0$ for any $x\ge0$, it is obvious that the right hand side of \eqref{eq26} is also nonnegative.

(2) For $\rho=\ket{\psi}\bra{\psi}$, the right hand sides of \eqref{eq24}-\eqref{eq26} reduce to
$$
\frac{1}{(\alpha-1)\beta}\Bigg\{\frac{\alpha-1}{\alpha}\mathrm{L}_{\frac{1}{\alpha}}\Bigg(\frac{(1-c)(d-1)}{MNS}+\frac{1}{N}\Bigg)+1\Bigg\}^{\alpha\beta}-\frac{1}{(\alpha-1)\beta},
$$

$$
-\frac{1}{\alpha\beta}\Bigg\{\frac{\alpha}{\alpha-1}\mathrm{L}_{\frac{1}{1-\alpha}}\Bigg(\frac{(1-c)(d-1)}{MNS}+\frac{1}{N}\Bigg)+1\Bigg\}^{(1-\alpha)\beta}+\frac{1}{\alpha\beta},
$$
and
$$
L_{(\alpha,\beta)}\Bigg(\frac{1}{\alpha\beta}\Bigg\{\frac{\alpha}{\alpha-1}\mathrm{L}_{\frac{1}{1-\alpha}}\Bigg(\frac{(1-c)(d-1)}{MNS}+\frac{1}{N}\Bigg)+1\Bigg\}^{(\alpha-1)\beta}-\frac{1}{\alpha\beta}\Bigg),
$$
respectively.

When the MUETFs in Theorem 1 reduce to MUBs and ETFs respectively,
we obtain the following two corollaries.

{\bf Corollary 1} Let $\mathcal{B} = \{\mathcal{B}_1, \mathcal{B}_2,
\cdots, \mathcal{B}_M\}$ be a set of MUBs in $\mathcal{H}$, and
$\rho\in\mathcal{D(H)}$. Then we have

(1) For any $\alpha\in[\frac{1}{2},1)$ and
$\beta\in(-\infty,0)\cup(0,1]$, it holds that
\begin{align}
\label{eq29}
\frac{1}{M}\sum\limits_{\mu=1}^{M}C_{(\alpha,\beta)}(\mathcal{B}_\mu;\rho)\ge&\frac{(\mathrm{tr}\rho^\alpha)^\beta}{(\alpha-1)\beta}\Bigg\{\frac{\alpha-1}{\alpha}\mathrm{L}_{\frac{1}{\alpha}}\Bigg(\frac{M-1+d\mathrm{tr}\rho^{2\alpha}(\mathrm{tr}\rho^\alpha)^{-2}}{Md}\Bigg)\notag\\
&+1\Bigg\}^{\alpha\beta}-\frac{1}{(\alpha-1)\beta};
\end{align}

(2) For any $\alpha\in(0,\frac{1}{2})$ and $\beta\in(-\infty,0)$, it
holds that
\begin{align}
\label{eq30}
\frac{1}{M}\sum\limits_{\mu=1}^{M}C_{(\alpha,\beta)}(\mathcal{B}_\mu;\rho)\ge&-\frac{(\mathrm{tr}\rho^{1-\alpha})^{\beta}}{\alpha\beta}\Bigg\{\frac{\alpha}{\alpha-1}\mathrm{L}_{\frac{1}{1-\alpha}}\Bigg(\frac{M-1+d\mathrm{tr}\rho^{2(1-\alpha)}(\mathrm{tr}\rho^{1-\alpha})^{-2}}{Md}\Bigg)\notag\\
&+1\Bigg\}^{(1-\alpha)\beta}+\frac{1}{\alpha\beta};
\end{align}

(3) For any $\alpha\in(0,\frac{1}{2})$ and $\beta\in(0,1]$, it holds
that
\begin{align}
\label{eq31}
\frac{1}{M}\sum\limits_{\mu=1}^{M}C_{(\alpha,\beta)}(\mathcal{B}_\mu;\rho)\ge&L_{(\alpha,\beta)}\Bigg(\frac{(\mathrm{tr}\rho^{1-\alpha})^{-\beta}}{\alpha\beta}\Bigg\{\frac{\alpha}{\alpha-1}\mathrm{L}_{\frac{1}{1-\alpha}}\Bigg(\frac{M-1+d\mathrm{tr}\rho^{2(1-\alpha)}(\mathrm{tr}\rho^{1-\alpha})^{-2}}{Md}\Bigg)\notag\\
&+1\Bigg\}^{(\alpha-1)\beta}-\frac{1}{\alpha\beta}\Bigg).
\end{align}

{\bf Remark 7} Letting $\alpha\to1$ and $\beta = 1$, Corollary 1 (1) reduces to Proposition 1 in \cite{Rastegin2018}. Letting $\alpha=\frac{1}{2}$ and $\beta = 1$, Corollary 1 (1) reduces to Theorem 1 in \cite{Rastegin2022}. Letting $\beta = 1$, Corollary 1 (1) reduces to partial results of Theorem 1 in \cite{Rastegin20241}, where the latter discusses the case for $\alpha\in[\frac{1}{2},1)\cup(1,+\infty)$.

{\bf Corollary 2} Let $\{\ket{\varphi_j}\}_{j=1}^{N}$ be an ETF in
$\mathcal{H}$, and $\rho\in\mathcal{D(H)}$. Then we have

(1) For any $\alpha\in[\frac{1}{2},1)$ and
$\beta\in(-\infty,0)\cup(0,1]$, it holds that
\begin{align}
\label{eq32}
C_{(\alpha,\beta)}(\mathcal{P};\rho)\ge&\frac{(\mathrm{tr}\rho^\alpha)^\beta}{(\alpha-1)\beta}\Bigg\{\frac{\alpha-1}{\alpha}\mathrm{L}_{\frac{1}{\alpha}}\Bigg(\frac{(1-c)[d\mathrm{tr}\rho^{2\alpha}(\mathrm{tr}\rho^\alpha)^{-2}-1]}{NS}\notag\\
&+\frac{1}{N}\Bigg)+1\Bigg\}^{\alpha\beta}-\frac{1}{(\alpha-1)\beta};
\end{align}

(2) For any $\alpha\in(0,\frac{1}{2})$ and $\beta\in(-\infty,0)$, it
holds that
\begin{align}
\label{eq33}
C_{(\alpha,\beta)}(\mathcal{P};\rho)\ge&-\frac{(\mathrm{tr}\rho^{1-\alpha})^{\beta}}{\alpha\beta}\Bigg\{\frac{\alpha}{\alpha-1}\mathrm{L}_{\frac{1}{1-\alpha}}\Bigg(\frac{(1-c)[d\mathrm{tr}\rho^{2(1-\alpha)}(\mathrm{tr}\rho^{1-\alpha})^{-2}-1]}{NS}\notag\\
&+\frac{1}{N}\Bigg)+1\Bigg\}^{(1-\alpha)\beta}+\frac{1}{\alpha\beta};
\end{align}

(3) For any $\alpha\in(0,\frac{1}{2})$ and $\beta\in(0,1]$, it holds
that
\begin{align}
\label{eq34}
C_{(\alpha,\beta)}(\mathcal{P};\rho)\ge&L_{(\alpha,\beta)}\Bigg(\frac{(\mathrm{tr}\rho^{1-\alpha})^{-\beta}}{\alpha\beta}\Bigg\{\frac{\alpha}{\alpha-1}\mathrm{L}_{\frac{1}{1-\alpha}}\Bigg(\frac{(1-c)[d\mathrm{tr}\rho^{2(1-\alpha)}(\mathrm{tr}\rho^{1-\alpha})^{-2}-1]}{NS}\notag\\
&+\frac{1}{N}\Bigg)+1\Bigg\}^{(\alpha-1)\beta}-\frac{1}{\alpha\beta}\Bigg),
\end{align}
where $\mathcal{P}$ is a POVM given in \eqref{eq15}, $c=\dfrac{N-d}{d(N-1)}$ and $S=\frac{N}{d}$.

{\bf Remark 8} Letting $\alpha=\frac{1}{2}$ and $\beta = 1$, Corollary 2 (1) reduces to Theorem 2 in \cite{Rastegin2022}. Letting $\beta = 1$, Corollary 2 (1) reduces to partial results of Theorem 2 in \cite{Rastegin20241}, where the latter discusses the case for $\alpha\in[\frac{1}{2},1)\cup(1,+\infty)$.

Letting $\beta=1$ in Theorem 1 (1) and (3), respectively, we obtain the following corollary, which gives the uncertainty relations via Tsallis $\alpha$-relative entropy of coherence $C_{\alpha}(\mathcal{F}_\mu;\rho)$.

{\bf Corollary 3} Let $\{\ket{\varphi_{\mu,j}}\}$ with
$\mu=1,\cdots,M$ and $j=1,\cdots,N$ be a MUETF in $\mathcal{H}$, and
$\rho\in\mathcal{D(H)}$. Then we have

(1) For any $\alpha\in[\frac{1}{2},1)$, it holds that
\begin{align}
\label{eq35}
\frac{1}{M}\sum\limits_{\mu=1}^{M}C_{\alpha}(\mathcal{F}_\mu;\rho)\ge&\frac{\mathrm{tr}\rho^\alpha}{\alpha-1}\Bigg\{\frac{\alpha-1}{\alpha}\mathrm{L}_{\frac{1}{\alpha}}\Bigg(\frac{(1-c)[d\mathrm{tr}\rho^{2\alpha}(\mathrm{tr}\rho^\alpha)^{-2}-1]}{MNS}\notag\\
&+\frac{1}{N}\Bigg)+1\Bigg\}^{\alpha}-\frac{1}{\alpha-1};
\end{align}

(2) For any $\alpha\in(0,\frac{1}{2})$, it holds that
\begin{align}
\label{eq36}
\frac{1}{M}\sum\limits_{\mu=1}^{M}C_{\alpha}(\mathcal{F}_\mu;\rho)\ge&\mathrm{L_{(\alpha,1)}}\Bigg(\frac{1}{\alpha\mathrm{tr}\rho^{1-\alpha}}\Bigg\{\frac{\alpha}{\alpha-1}\mathrm{L}_{\frac{1}{1-\alpha}}\Bigg(\frac{(1-c)[d\mathrm{tr}\rho^{2(1-\alpha)}(\mathrm{tr}\rho^{(1-\alpha)})^{-2}-1]}{MNS}\notag\\
&+\frac{1}{N}\Bigg)+1\Bigg\}^{\alpha-1}-\frac{1}{\alpha}\bigg),
\end{align}
where $\mathcal{F_\mu}$ are the induced POVMs given in \eqref{eq20}, $c=\dfrac{N-d}{d(N-1)}$ and $S=\frac{N}{d}$.

{\bf Remark 9} Corollary 3 (1) is a partial result of Proposition 1 in \cite{Rastegin20242}, where the latter discusses the case for $\alpha\in[\frac{1}{2},1)\cup(1,+\infty)$.

Letting $\beta\to0$ in Theorem 1 (1) and in Theorem 1 (2)/(3),
respectively, we obtain the first and second item of the following
corollary, which are the uncertainty relations via Re\'nyi
$\alpha$-relative entropy of coherence
$\widetilde{C}_{\alpha}(\mathcal{F}_\mu;\rho)$.

{\bf Corollary 4} Let $\{\ket{\varphi_{\mu,j}}\}$ with
$\mu=1,\cdots,M$ and $j=1,\cdots,N$ be a MUETF in $\mathcal{H}$, and
$\rho\in\mathcal{D(H)}$. Then we have

(1) For any $\alpha\in[\frac{1}{2},1)$, it holds that
\begin{align}
\label{eq37}
\frac{1}{M}\sum\limits_{\mu=1}^{M}\widetilde{C}_{\alpha}(\mathcal{F}_\mu;\rho)\ge&\frac{1}{\alpha-1}\mathrm{ln}\Bigg(\Bigg\{\frac{\alpha-1}{\alpha}\mathrm{L}_{\frac{1}{\alpha}}\Bigg(\frac{(1-c)[d\mathrm{tr}\rho^{2\alpha}(\mathrm{tr}\rho^\alpha)^{-2}-1]}{MNS}\notag\\
&+\frac{1}{N}\Bigg)+1\Bigg\}^{\alpha}\mathrm{tr}\rho^\alpha\bigg);
\end{align}

(2) For any $\alpha\in(0,\frac{1}{2})$, it holds that
\begin{align}
\label{eq38}
\frac{1}{M}\sum\limits_{\mu=1}^{M}\widetilde{C}_{\alpha}(\mathcal{F}_\mu;\rho)(\mathcal{F}_\mu;\rho)\ge&-\frac{1}{\alpha}\mathrm{ln}\Bigg(\Bigg\{\frac{\alpha}{\alpha-1}\mathrm{L}_{\frac{1}{1-\alpha}}\Bigg(\frac{(1-c)[d\mathrm{tr}\rho^{2(1-\alpha)}(\mathrm{tr}\rho^{(1-\alpha)})^{-2}-1]}{MNS}\notag\\
&+\frac{1}{N}\Bigg)+1\Bigg\}^{(1-\alpha)}\mathrm{tr}\rho^{(1-\alpha)}\bigg),
\end{align}
where $\mathcal{F_\mu}$ are the induced POVMs given in \eqref{eq20}, $c=\dfrac{N-d}{d(N-1)}$ and $S=\frac{N}{d}$.

\noindent{\bf 4 Examples}\par

To exemplify the obtained results, we consider the following examples.

{\bf Example 1} For any $\alpha\in[\frac{1}{2},1)$, let
$\beta=\alpha$, $d=2$, $N=2$ and $M=3$, $\mathcal{B} =
\{\mathcal{B}_1, \mathcal{B}_2, \mathcal{B}_3\}$ be a set of MUBs
with
$\mathcal{B}_1=\left\{\frac{\ket{0}+\ket{1}}{\sqrt{2}},\frac{\ket{0}-\ket{1}}{\sqrt{2}}\right\}$,
$\mathcal{B}_2=\left\{\frac{\ket{0}+\mathrm{i}\ket{1}}{\sqrt{2}},\frac{\ket{0}-\mathrm{i}\ket{1}}{\sqrt{2}}\right\}$,
$\mathcal{B}_3=\left\{\ket{0},\ket{1}\right\}$. Consider the
pseudopure states
$$
\rho=\frac{1-v}{2}\mathbf{I_2}+v\ket{0}\bra{0},
$$
where $v\in[0,1]$ and $\mathbf{I_2}$ is the $2\times 2$ identity
matrix. Direct calculations show that the left and right hand side
of \eqref{eq29} becomes
\begin{align}
\label{eq39}
\frac{2-2^{1-\alpha } [(1-v)^{\alpha }+(v+1)^{\alpha }]^{\alpha}}{3 \alpha -3 \alpha ^2}
\end{align}
and
\begin{align}
\label{eq40} &\frac{((1 - v)^\alpha + (1 +
v)^\alpha)^\alpha}{(\alpha-1)\alpha2^{\alpha^2}}&\Bigg\{\frac{\alpha-1}{\alpha}\mathrm{L}_{\frac{1}{\alpha}}\Bigg(
\frac{2 ((1-v^2)^{\alpha }+(1-v)^{2 \alpha }+(v+1)^{2 \alpha })}{3
((1-v)^{\alpha }+(v+1)^{\alpha })^2} \Bigg)
+1\Bigg\}^{\alpha^2}\notag\\
&-\frac{1}{(\alpha-1)\alpha},
\end{align}
respectively.

Figure 1 presents the coherence quantifier averaged over the three
MUBs in $\mathcal{H}_2$ for pseudopure state and the corresponding
lower bound, while Figure 2 depicts the gap between them as a
function of $v$ for fixed parameters $\alpha$ and as a function of
$\alpha$ for fixed parameters $v$.

\begin{figure}[H]\centering
\includegraphics[width=0.5\textwidth]{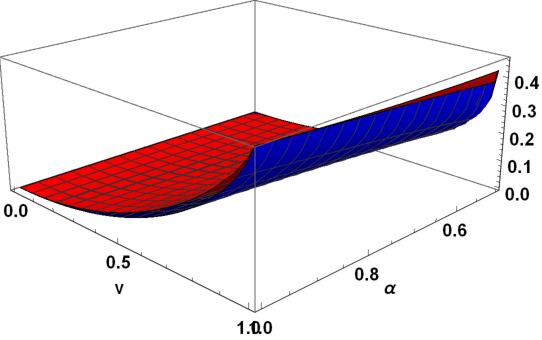}
\caption{{Uncertainty relations via unified-($\alpha$,$\beta$)
relative entropy with $\beta=\alpha\in[\frac{1}{2},1)$ under three
MUBs. The red surface represents the quantity in \eqref{eq39}, and
the blue surface represents the quantity in \eqref{eq40}.
\label{fig:Fig1}}}
\end{figure}

\begin{figure}[H]
    \centering
    \begin{minipage}[b]{0.48\textwidth}
        \centering
        \includegraphics[width=\linewidth]{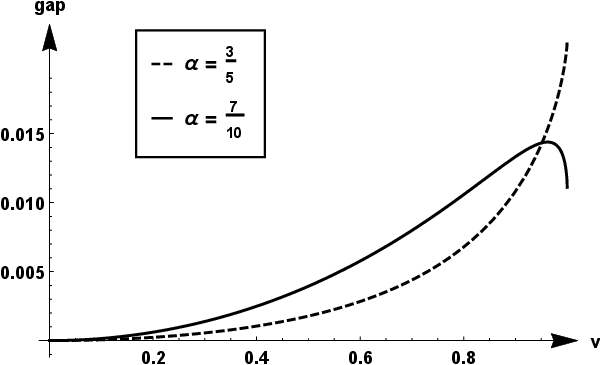} 
        \textbf{(a)} 
        \label{fig:Fig2a}
    \end{minipage}
    \hfill
    \begin{minipage}[b]{0.48\textwidth}
        \centering
        \includegraphics[width=\linewidth]{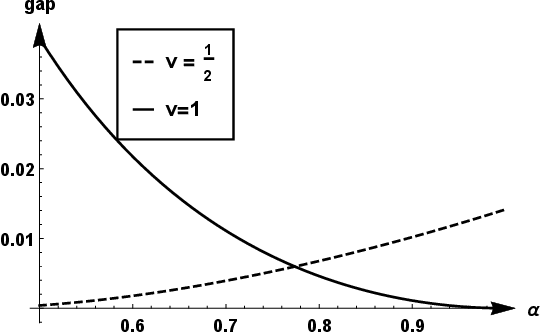} 
        \textbf{(b)}
        \label{fig:Fig2b}
    \end{minipage}
    \caption{Curves of the gap between \eqref{eq39} and \eqref{eq40} with fixed $\alpha$ and $v$: (a) $\alpha=\frac{3}{5}$ and $\alpha=\frac{7}{10}$; (b) $v=\frac{1}{2}$ (pseudopure state) and $v=1$ (pure state).}
    \label{fig:Fig2}
\end{figure}

It is shown that the range of variations is so narrow here and does
not exceed 0.1, which demonstrates that the lower bounds give a good
estimation of the average coherence in this specific case. The
average coherence and the corresponding lower bounds are both convex
and increasing with respect to $v$ for fixed $\alpha$, and with
respect to $\alpha$ for fixed $v$. Numerical calculations show that
for fixed $\alpha$, the gap between \eqref{eq39} and \eqref{eq40}
becomes larger or larger first and smaller then when $v$ is larger,
depending on the value of $\alpha$, while for fixed $v$, the gap
between \eqref{eq39} and \eqref{eq40} may be larger or smaller when
$\alpha$ is larger.

{\bf Example 2} For any $\alpha\in[\frac{1}{2},1)$, set
$\beta=-\alpha$, $d=2$, $N=4$ and $M=1$. A qubit state is in the
form $\rho=\frac{\mathbf{I_2}+\overrightarrow{r} \cdot
\overrightarrow{\sigma}}{2}$, where $\mathbf{I_2}$ is the $2\times
2$ identity matrix, $\overrightarrow{r}=(r_1,r_2,r_3)$ is the Bloch
vector and $\overrightarrow{\sigma}=(\sigma_x,\sigma_y,\sigma_z)$ is
composed of Pauli matrices. Now assume that
$\overrightarrow{r}=(r_1,0,r_1)$. Consider the SIC-POVM $\mathcal{F}
= \{ \frac{1}{2}\ket{\phi_i}\bra{\phi_i} \}_{i=0}^{3}$ with
\cite{Rastegin20232}
$$
\ket{\varphi_0} = \ket{0}, \ket{\varphi_1} = \frac{1}{\sqrt{3}}(\ket{0}+\sqrt{2}\ket{1}), \ket{\varphi_2} =\frac{1}{\sqrt{3}}( \ket{0}+\sqrt{2}\omega\ket{1}), \ket{\varphi_3} =\frac{1}{\sqrt{3}}( \ket{0}+\sqrt{2}\omega^*\ket{1}),
$$
where $\omega = e^{\frac{2\pi\mathrm{i}}{3}}$. Since in this case $N=d^2$, the above SIC-POVM is an ETF.

Direct calculations show that the left and right hand side of
\eqref{eq32} becomes
\begin{align}
\label{eq41} &\frac{1}{(1-\alpha) \alpha}((12^{-\frac{1}{\alpha}}
(3^{\frac{1}{\alpha}} (2^{-\alpha-\frac{1}{2}} ((\sqrt{2}+1)
(\sqrt{2} r_1+1)^{\alpha}+(\sqrt{2}-1) (1-\sqrt{2}
r_1)^{\alpha}))^{\frac{1}{\alpha}}\notag\\&+ 2
(2^{-\alpha-\frac{1}{2}} ((2 \sqrt{2}-1)(\sqrt{2} r_1+1)^{\alpha}+(4
\sqrt{2}+1) (1-\sqrt{2}
r_1)^{\alpha}))^{\frac{1}{\alpha}}\notag\\&+(2^{-\alpha-\frac{1}{2}}
((5 \sqrt{2}-1) (\sqrt{2} r_1+1)^{\alpha}+(\sqrt{2}+1) (1-\sqrt{2}
r_1)^{\alpha}))^{\frac{1}{\alpha}}))^{-\alpha^2}-1).
\end{align}
and
\begin{align}
\label{eq42}
\frac{2^{\alpha^2}\Bigg\{\frac{\alpha-1}{\alpha}\mathrm{L}_{\frac{1}{\alpha}}\Bigg( \frac{\left(1-2 r_1^2\right)^\alpha+\left(\sqrt{2} r_1+1\right)^{2 \alpha}+\left(1-\sqrt{2} r_1\right)^{2 \alpha}}{3 \left(\left(\sqrt{2} r_1+1\right)^\alpha+\left(1-\sqrt{2} r_1\right)^\alpha\right)^2} \Bigg)+1\Bigg\}^{-\alpha^2}}{(1-\alpha)\alpha((\sqrt{2} r_1+1)^\alpha+(1-\sqrt{2} r_1)^\alpha)^\alpha}-\frac{1}{(1-\alpha)\alpha},
\end{align}
respectively.

Figure 3 shows the coherence quantifier averaged over the SIC-POVMs
in $\mathcal{H}_2$ for a qubit state and the corresponding lower
bound, and Figure 4 depicts the gap between them for fixed $\alpha$
and fixed $r_1$.

\begin{figure}[H]\centering
\includegraphics[width=0.5\textwidth]{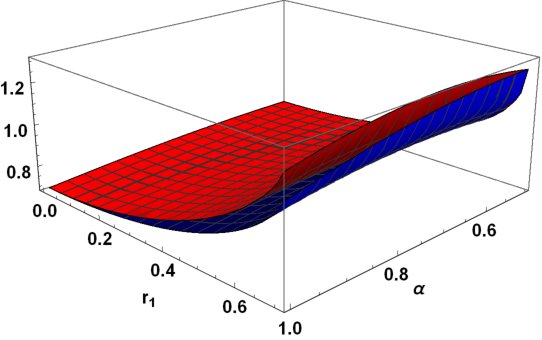}
\caption{{Uncertainty relations via unified-($\alpha$,$\beta$) relative entropy with $-\beta=\alpha\in[\frac{1}{2},1)$ under a set of SIC-POVMs. The red surface represents the quantity in \eqref{eq41}, and the blue surface represents the quantity in \eqref{eq42}. \label{fig:Fig3}}}
\end{figure}

\begin{figure}[H]
    \centering
    \begin{minipage}[b]{0.48\textwidth}
        \centering
        \includegraphics[width=\linewidth]{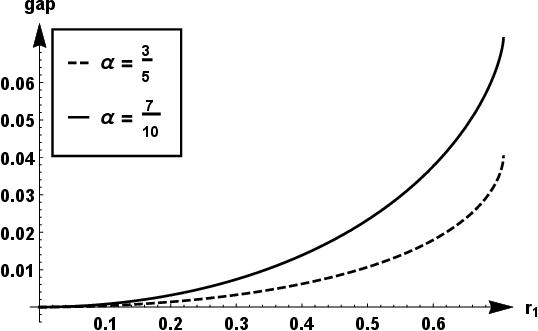} 
        \textbf{(a)} 
        \label{fig:Fig4a}
    \end{minipage}
    \hfill
    \begin{minipage}[b]{0.48\textwidth}
        \centering
        \includegraphics[width=\linewidth]{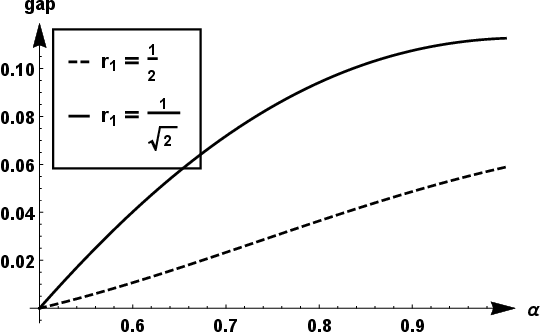} 
        \textbf{(b)}
        \label{fig:Fig4b}
    \end{minipage}
    \caption{Curves of the gap between \eqref{eq41} and \eqref{eq42} with fixed $\alpha$ and $r_1$: (a) $\alpha=\frac{3}{5}$ and $\alpha=\frac{7}{10}$; (b) $r_1=\frac{1}{2}$ (mixed state) and $r_1=\frac{1}{\sqrt{2}}$ (pure state).}
    \label{fig:Fig4}
\end{figure}

It is demonstrated that the averaged coherence quantifiers and the
corresponding lower bounds are both convex with respect to $r_1$ for
fixed $\alpha$, the former increases first and then decreases with
respect to $\alpha$, while the latter decreases with respect to
$\alpha$ for fixed $r_1$, and closely adheres to the corresponding
surface of averaged coherence during the change process. Numerical
calculations show that for fixed $\alpha$, the gap between
\eqref{eq41} and \eqref{eq42} becomes larger when $r_1$ is larger,
and for fixed $r_1$, this gap also becomes larger when $\alpha$ is
larger. The range of variations between the averaged quantifiers and
the corresponding lower bounds is so narrow that the latter can be
seen as a good approximation of the former under this circumstance.

\vskip0.1in

\noindent {\bf 5. Conclusions}\par Using the unified
($\alpha$,$\beta$)-relative entropy of coherence, the uncertainty
relations for the quantifiers averaged over POVMs assigned to
MUETFs, which are state-dependent, has been derived. The
inequalities offered a unified approach to quantify uncertainty of
coherence, making it applicable to a broad range of quantum
information tasks. In specific circumstances, the unified
($\alpha$,$\beta$)-relative entropy of coherence reduce to special
coherence quantifiers, and MUETFs reduce to MUBs or ETFs, so our
results are natural generalizations of the results in previous
literatures. The inequalities has been illustrated using SIC-POVMs
and MUBs in two dimensional spaces, indicating that the lower bound
provides a good approximation in some situations. The results in
this paper may shed some new light on the research of uncertainty
relations based on coherence quantifiers under a set of bases or
measurements. Note that if the state $\sigma$ in \eqref{eq8} is
invertible, then the definition of unified
($\alpha$,$\beta$)-relative entropy can be extended to $\alpha>1$
\cite{WangJ}. In this case, if there exisits $\beta\in\mathbb{R}$,
such that the function $C_{(\alpha,\beta)}(\mathcal{A};\rho)$ in
\eqref{eq10} is a coherence monotone, we can further discuss the
uncertainty relations for a broader range of parameters. This is
left for further study.

\vskip0.1in

\noindent

\subsubsection*{Acknowledgements}
The authors would like to express their sincere gratitude to the
anonymous referees for their suggestions, which greatly improved the
paper. This work was supported by National Natural Science
Foundation of China (Grant No. 12161056) and Natural Science
Foundation of Jiangxi Province of China (Grant No. 20232ACB211003).

\subsubsection*{Author Contributions}
Baolong Cheng wrote the main manuscript text and Zhaoqi Wu
supervised and revised the manuscript. All authors reviewed the
manuscript.

\subsubsection*{Data Availability}
No datasets were generated or analysed during the current study.

\subsubsection*{Competing interests}
The authors declare no competing interests.

\end{document}